\documentstyle[12pt]{article}
\begin{document}
\begin{center}
{\bf Locality Is An Unnecessary Assumption of Bell's Theorem}\\
\vspace{0.5cm}

{\bf H. Razmi}\\

\vspace{0.5cm}

{\it \small Department of Physics, School of sciences,\\
Tarbiat Modarres University, P.O.Box 14155-4838\\
Tehran, I.R. Iran }\\
{\bf M.Golshani}\\
\vspace{0.5cm}
{\it \small Department of Physics, Sharif University of Technology,\\
P.O.Box 11365-9161,Tehran,I.R.Iran}\\
{\it \small Institute for Studies in Theoretical Physics and Mathematics,\\
P.O.Box 19395-1795,Tehran,I.R.Iran}\\
{\bf Abstract}\\
\end{center}
Without imposing the locality condition, it is shown that quantum theory cannot
reproduce all the predictions of a special stochastic realistic model in certain
spin-correlation experiments.This shows that the so-called locality condition
is an unnecessary assumption of Bell's theorem.\\

\newpage
Locality condition has an important role in many models of Bell's theorem.
In Bell's original work [1], CH model [2], and almost all the other related
works, a condition often used to derive different forms of Bell's inequality
is the so-called locality condition.Here we are going to prove Bell's theorem
based on a nonlocal model (i.e. a model in which we don't impose the locality 
condition).In fact,we are going to show that quantum mechanics cannot reproduce
all the predictions of a (the following) stochastic realistic model in certain spin-correlation
experiments.\\  
Consider a source which emits a pair of spin $\frac{1}{2}$ identical particles
in a singlet state (i.e. with total angular momentum).The two particles of the
pair travel in opposite directions toward suitable measuring devices $D_A$ and 
$D_B$.
Identify each pair of escaping particles by an index i (i=1,2,...,n).The
experimental arrangement is such that one particle from each pair will enter
$D_A$, and the other one will enter $D_B$.\\
Each detector has a preferred direction.The azimuthal orientation of this
direction denoted by $Q_A$ and $Q_B$ according to whether one deals with ${\theta}_A$
or ${\theta}_B$.Denote by $r_{ji}$ the result obtained by $D_j$ in the ith pair.
This can take values +1 or -1 according to whether the deflection is along the
preferred direction or its opposite.Each detector can take two possible values
${\theta}_j'$ and ${\theta}_j''$.A crucial requirement is that the 
experimental set-up in sides A and B be spatially separated (i.e. having
space-like separation in the sense of the special theory of relativity).\\
There are four possible experimental specifications corresponding to the four
possible pairs $({\theta}_A',{\theta}_B')$ , $({\theta}_A',{\theta}_B'')$
, $({\theta}_A'',{\theta}_B')$ and $({\theta}_A'',{\theta}_B'')$.\\
By definition the correlation function $C(r_A,r_B)$ is the average of the
product of the results obtained by $D_A$ and $D_B$ :\\
\begin{equation}
C(r_A,r_B) = \frac{1}{n}\sum_{i=1}^n r_{Ai}({\theta}_A,{\theta}_B)r_{Bi}({\theta}_A,{\theta}_B)
\end{equation}
People$^{\dag}$ [3-4] usually impose the following locality condition:\\
\begin{equation}
\left\{\begin{array}{lll}
r_A({\theta}_A,{\theta}_B) = r_A({\theta}_A)\\
r_B({\theta}_A,{\theta}_B) = r_B({\theta}_B)
\end{array}\right.
\end{equation}
and write the correlation function (1) in the following form:\\
\begin{equation}
C(r_A,r_B) = \frac{1}{n}\sum_{i=1}^n r_{Ai}({\theta}_A)r_{Bi}({\theta}_B)
\end{equation}
and then try to find an inequality (Bell's inequality) which cannot be always
satisfied by quantum mechanics.\\
Here we are going to show that even before the introduction of the locality condition
(2) (i.e. only working with the correlation function (1)),there is a descrepancy
between the model used and quantum theory.\\
We know that the quantum mechanical result for the correlation function of the
above-mentioned experimental set-up (the singlet state) is:\\
\begin{equation}
C_{QM}(r_A,r_B) = -\cos({\theta}_A-{\theta}_B)
\end{equation}
Now,let equate the correlation functions (1) and (4):\\
\begin{equation}
\frac{1}{n}\sum_{i=1}^n r_{Ai}({\theta}_A,{\theta}_B)r_{Bi}({\theta}_A,{\theta}_B) = -\cos({\theta}_A-{\theta}_B)
\end{equation}
Since both $r_{Ai}$ and $r_{Bi}$ are dichotomic variables (i.e. take values +1 or -1 only),their product 
is also dichotomic.Assume that from n runs of the experiment,m of them
show the value +1 for $r_Ar_B$;thus,n-m of them would show the value -1.Therefore,
the relation (5) is reduced to the following form:\\
\begin{equation}
\frac{1}{n}([\sum_{i=+1}^m (+1)] + [\sum_{i=m+1}^n (-1)]) = -\cos({\theta}_A-{\theta}_B)
\end{equation}
Or:\\
\begin{equation}
\frac{m}{n} = {\sin}^2(\frac{{\theta}_A-{\theta}_B}{2})
\end{equation}
For arbitrary values of ${\theta}_A$ and ${\theta}_B$, the right hand side of
(7) is not necessarily a rational number.On the other hand m and n are,by definition,
integers.Therefore,$\frac{m}{n}$ is always a rational number.Thus,the relation
(7) cannot generally hold.Of course, for very large values of n,it seems there 
is no problem.But,we can show that in the actual experiments the relation (7)
could be noticeably violated.To illustrate this,we appeal to the data of one of
the actual experiments.In the proton-proton scatering experiment of Lamehi-Rachti
and Mittig [5],which exemplifies the above set-up,the total count was $10^4$ [6].
Now, if we take:\\
$${\theta}_A = 47.4^{\circ} , {\theta}_B = 45^{\circ}$$\\
Then,\\
$${\sin}^2(\frac{{\theta}_A-{\theta}_B}{2}) = 4.4\times10^{-4}$$\\
Thus:\\
$$\frac{m}{n} = 4.4\times10^{-4}$$\\
Or,\\
$$m = 4.4$$\\
Since the nearest integer to $4.4$ is 4, the relative error in making m an integer
is $\frac{0.4}{4}$ or $10^{-1}$,which isn't an ignorable,negligible, value.\\
Therefore,the quantum mechanical correlation function (4) cannot be equated
to the correlation function (1) for all possible values of ${\theta}_A$ and 
${\theta}_B$.This shows that even before the introduction of the locality 
condition (2), quantum theory cannot reproduce all the predictions of the
stochastic realistic model used.\\
In conclusion,locality condition is not a necessary assumption to show that 
quantum theory is in conflict with some of the predictions of certain stochastic
realistic models (i.e.to prove Bell's theorem).
 
\newpage 

\begin{center}
{\bf REFERENCES} 
\end{center}

\begin{itemize}
\item[1] J.S.Bell,Physics {\bf 1},No.3,195 (1964).
\item[2] J.F.Clauser and M.A.Horne,Phys.Rev.D {\bf 10},526 (1974).
\item[${\dag}$] Although the model used here is equivalent 
,even in the notation, to the Stapp's model [3-4];since we haven't 
used the locality condition here,there is an essential difference
between our work and Stapp's model.
\item[3] H.P.Stapp,Am.J.Phys. {\bf 53},306 (1985).
\item[4] H.P.Stapp,Found.Phys. {\bf 18},427 (1988).
\item[5] M.Lamehi-Rachti and W.Mittig,Phys.Rev. {\bf 14},2543 (1976).
\item[6] Private Communication with Dr.Lamehi-Rachti.
\end{itemize}

\end{document}